\documentclass[twoside]{article}
\usepackage{fleqn,espcrc2,epsfig}

\usepackage{graphicx}

\newcommand{\be}{\begin{equation}}
\newcommand{\ee}{\end{equation}}

\title{Laplacian gauge and instantons}

\author{Philippe de Forcrand\address[ETH]{Inst. f\"ur Theoretische
    Physik, ETH H\"onggerberg, CH-8093 Z\"urich, Switzerland}
\address{CERN, Theory Division, CH-1211 Gen\`eve 23, Switzerland}
\thanks{Presented by Ph. de Forcrand at LAT00.}
  and Michele Pepe\addressmark[ETH]}

\begin{document}

\begin{abstract}
We exhibit the connection between local gauge singularities in the Laplacian 
gauge and topological charge, which opens the possibility of studying
instanton excitations without cooling. 
We describe our version of Laplacian gauge-fixing for $SU(N)$.
\vspace{-0.4cm}
\end{abstract}

\maketitle

Topological excitations produce obstructions to making the gauge field smooth
everywhere. Therefore, they should appear as singularities in an otherwise
smooth gauge. This offers the possibility of identifying such excitations
via gauge-fixing. After gauge-fixing, the gauge field becomes singular,
and the gauge ill-defined, on a sub-manifold characterizing the topological
excitations. Even though the precise location of this manifold typically 
depends on the specific gauge condition chosen, its existence does not.
This gauge-fixing approach was suggested by 't Hooft to identify 
chromo--magnetic monopoles \cite{tH}. It has been recognized recently that
both monopoles and center vortices appear together as gauge singularities
of co-dimension 3 and 2 respectively, when one tries to enforce a smooth
gauge for the adjoint $SU(N)/Z_N$ field \cite{CV}. It is then natural to
also study what happens when one tries to enforce a smooth gauge for the
$SU(N)$ field. As we show below, point-like (co-dimension 4) singularities
appear, coming from the topological charge of the Yang-Mills field.
Thus, gauge-fixing allows a unifying perspective on all 3 kinds of topological
Yang-Mills excitations: center vortices, monopoles and instantons.

\section{Example: Landau gauge}

\noindent
The gauge field of an $SU(2)$ instanton of size $\rho$ is
\be
A_\mu^a(x) = \eta_{\mu\nu}^a x_\nu \frac{2}{|x|^2 + \rho^2} ~~{\rm or}~~
{\bar \eta_{\mu\nu}^a} x_\nu \frac{2 \rho^2}{|x|^2(|x|^2 + \rho^2)}~~
\ee
in regular and singular gauge respectively. Because of the antisymmetry of
$\eta_{\mu\nu}$ and ${\bar \eta_{\mu\nu}}$, both expressions above satisfy
the Landau gauge condition $\partial_\mu A_\mu = 0$. Furthermore,
starting from the singular gauge, one can apply the gauge transformation
$\Omega(x) = \omega(x,b) \omega(x,0)^\dagger$, where 
$\omega(x,y) = ((x-y)_0 {\bf 1} + i (x-y)_i \sigma_i)/|x-y|$,
and still obtain a configuration which satisfies $\partial_\mu A_\mu = 0$.
In this configuration, the gauge field is singular at point $b$, which can
be anywhere in 4-space. The regular gauge is just a special case where $b$
is sent to infinity. Therefore, Landau gauge is ambiguous on an instanton
background field, and a whole family of Gribov copies exists. Nevertheless,
the gauge field possesses a point-like singularity in each of these copies,
whose location depends on the copy chosen. In the presence of small
perturbations, the Gribov ambiguity would be resolved, but the location of
the singularity would be dictated by the perturbation, and would have a priori
little to do with the instanton center.

\section{Laplacian gauge: $SU(2)$}

On the lattice, the Landau gauge condition is usually translated into
\be
\sum_{x,\mu} ReTr~U_{x,\mu} ~~{\rm maximum}
\ee
which is ambiguous for any field configuration, due to the presence of
local maxima called ``lattice Gribov copies''. In a gauge given by
a local maximum, spurious singularities of the gauge field occur after
continuum interpolation,
which for instance bias the correlator $\langle A_\mu(0) A_\mu(x) \rangle$
\cite{Bielefeld}. Therefore, it is desirable to select a smooth gauge devoid
of lattice Gribov copies to study gauge field singularities. The Laplacian
gauge \cite{VW} is such a gauge.
Like Landau gauge, it is smooth and Lorentz-symmetric, and fixes a pure gauge
configuration ($F_{\mu\nu}=0$) to $A_\mu=0$ (all links $={\bf 1}$).
Our computer implementation for $SU(2)$ and $SU(3)$ uses the
library ARPACK \cite{ARPACK}, and
is much more economical than the iterative Landau 
gauge fixing. With it, a study of the infrared behaviour of the gluon propagator 
has been performed \cite{Dina}. 
Unfortunately, the Laplacian gauge is intrinsically non-perturbative,
which prevents a direct \linebreak comparison with the Faddeev-Popov approach.

The Laplacian gauge condition consists of imposing, at every space-time point,
a fixed color orientation for the eigenvector of the covariant Laplacian 
associated with the smallest eigenvalue. In other words, one determines
the eigenvector $v(x)$ satisfying
$\Delta(U)~v(x) = \lambda_0 v(x)$.
At each point $x$, $v(x) \in C^2$ has 2 complex color components.
The required gauge transformation $\Omega(x)$ is that which rotates $v(x)$
along some prescribed direction, say $(1,0)^\dagger$. 
The gauge thus specified is unique,
except when the eigenvalue $\lambda_0$ is degenerate, which defines a genuine
Gribov copy.\footnote{An irrelevant degeneracy 
$\vec{v} \leftrightarrow \sigma_2 \vec{v}^\star$ is always present.}

For a generic gauge field, the gauge may still be ill-defined 
{\em locally}:
this happens at points $x$ where $\Omega(x)$ cannot be defined because
$|v(x)| = 0$. This implies 4 constraints: the sub-manifold
where the gauge remains ambiguous consists of points
in 4 dimensions. It is natural to examine the connection between such points
and instantons.

\begin{figure}[t]
\begin{center}
\epsfig{figure=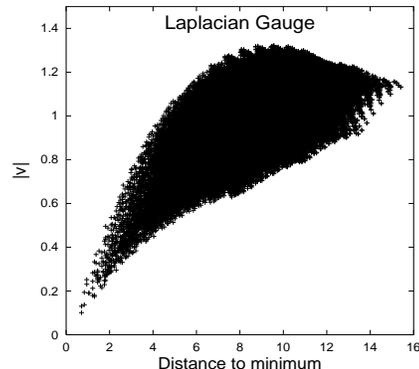,height=5.0cm,width=5.5cm}
\end{center}
\vskip -1.5cm
\caption{Magnitude of the lowest-lying eigenvector of the
covariant Laplacian, for a cooled instanton gauge field.}
\label{fig1}
\vskip -0.7cm
\end{figure}

Fig.1 shows $|v(x)|$ on an instanton configuration obtained by improved cooling
\cite{impcool}.
$|v(x)|$ appears to vanish at a point which turns out to
be very near the instanton center, and varies linearly with distance in its
vicinity. This behaviour has been shown analytically, for the continuum theory 
in a finite volume, in \cite{Wipf}. Thus, one instanton can be identified via 
Laplacian gauge-fixing.

\section{Many instantons}

In the field of $Q$ instantons, one expects $Q$ gauge singularities 
in Laplacian gauge.
Their locations cannot be obtained analytically, so we have performed
numerical tests on configurations of various topological content, obtained
by improved cooling. To identify the location of the instantons, we use
the ``instanton finder'' algorithm of \cite{impcool}, which looks for
a local peak in the topological charge density, and checks self-duality
near the peak. To identify the gauge singularity where the magnitude $|v(x)|$
of the Laplacian eigenvector vanishes, we fit $|v(x)|$ to the ansatz
\be
|v(x)| \sim v_0 + |x - x_0| / r_0
\label{ansatz}
\ee
on a $3^4$ hypercube. If the fit is good ($\chi^2/{\rm d.o.f.} \ll 1$, typically
$\le 0.004$) and $|v_0|$ is small (typically $\le 0.25$), we assign a
singularity at point $x_0$. Neighbouring $3^4$ cells usually detect the
same singularity, providing a cluster of estimates $x_0$.

We find very good agreement between the locations of the topological excitations
given by both methods. The agreement improves with cooling, and is worse on
approximate instantons which depart more from a spherical ansatz. Even
on a 7-instanton configuration, the agreement remains rather good, as shown
in Fig.2. 

\begin{figure}[t]
\begin{center}
\epsfig{figure=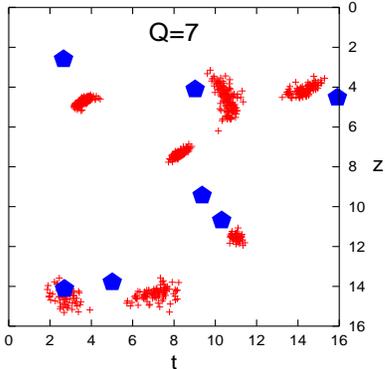,height=5.0cm,width=5.3cm}
\end{center}
\vskip -1.5cm
\caption{On a configuration containing 7 cooled instantons,
comparison of their locations (pentagons) and the approximate
zeros of the Laplacian eigenvector ($+$), 
projected on the $(z,t)$ plane.}
\label{fig2}
\vskip -0.9cm
\end{figure}

When both instantons and anti-instantons are present, the agreement between
the two methods deteriorates. The Laplacian method pays more attention to
the global features of the gauge field, which all enter in the construction
of the lowest-lying eigenvector, and does not always ``see'' an  
instanton-antiinstanton pair. The constant $v_0$ in the ansatz eq.(\ref{ansatz})
helps towards accounting for the background of the opposite charge excitations.
It is also possible to select an eigenvector of the Laplacian associated
with a larger eigenvalue, thus more sensitive to short-range gauge field 
fluctuations. The agreement with the usual method improves noticeably.

Similarly, one may try to analyze the topological content of a thermalized
gauge field configuration. The usual method requires some cooling, which
has been criticized because it destroys close instanton-antiinstanton pairs
and distorts the remaining objects. In contrast, the Laplacian method
filters the UV fluctuations {\em automatically}, so that information can
be obtained {\em without any cooling}. This is illustrated in Fig.3.
It is readily apparent that the two methods agree poorly in such circumstances.

\begin{figure}[th]
\begin{center}
\epsfig{figure=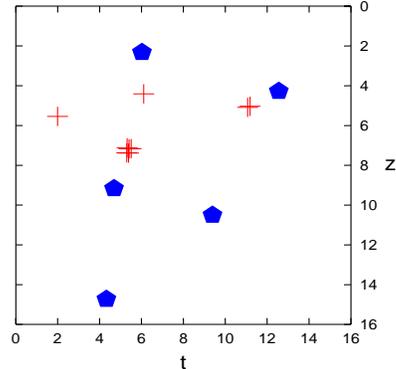,height=5.0cm,width=5.3cm}
\end{center}
\vskip -1.5cm
\caption{On a $\beta=2.4$ configuration,
comparison of the locations of approximate instantons (pentagons) 
and of approximate zeros of the Laplacian eigenvector ($+$), 
projected on the $(z,t)$ plane. The former required 30 cooling steps,
the latter zero.}
\label{fig3}
\vskip -0.8cm
\end{figure}

This difference emphasizes once more the arbitrariness which necessarily
enters any procedure which aims at identifying instantons and anti-instantons
in a fluctuating ensemble. Strictly speaking, there is no such thing as
an instanton-antiinstanton pair, since it does not constitute a solution
of the classical Yang-Mills equations. To distinguish such a pair from a
topologically trivial fluctuation, one must separate non-perturbative and
perturbative contributions by some criterion. This is a notoriously ambiguous 
proposition.

Therefore, one has here two methods, each containing some arbitrariness, to
identify topological excitations. The usual method analyzes {\em local} 
features of the gauge field, and requires some arbitrary amount of cooling. 
The Laplacian method takes into account {\em global} features through the
eigenvector. Which eigenvector one selects is arbitrary. In addition, both
methods encapsulate further arbitrariness in the quality of the local fit 
required, in both cases, to identify an instanton.

One possible advantage we see to the Laplacian method is in the measurement of 
the total topological charge. Small instantons are easily destroyed in the
early stage of cooling. The Laplacian method might give more reliable results.
Of course the near-vanishing of $|v(x)|$ gives no indication on the sign
of the associated topological charge, or the size and color orientation of
the instanton. One must look at the gauge-fixed field to extract such
information.

\section{Laplacian gauge: $SU(N)$}

A formulation of the Laplacian gauge for $SU(N)$ was given in \cite{VW}.
It requires the determination of the $N$ lowest-lying eigenvectors of the
covariant Laplacian. At each point $x$, the $N \times N$ complex matrix $M$ 
whose columns are formed by these eigenvectors is projected onto $SU(N)$, using
the polar decomposition $M = W P, W \in U(N), P \linebreak = (M^\dagger M)^{1/2}$. 
The required gauge transformation is then 
$\Omega(x) = e^{i \alpha} W^\dagger$, where 
$\alpha = \frac{1}{N} {\rm arg}(\det W)$.
$\Omega(x)$ rotates $M$ ``parallel'' to the identity ${\bf 1}_N$.

Here we use a different $SU(N)$ formulation, which requires $(N-1)$ 
eigenvectors only. Moreover, in this formulation, the connection between 
topological charge and local gauge ambiguities remains transparent. We consider 
the $(N-1)$ eigenvectors $v_1, v_2,..,v_{N-1}$ in succession, and associate
a partial gauge-fixing with each. \\
$i)$ At each point $x$, rotate $v_1(x)$ parallel to $(1,0,..,0)_N^\dagger$.
The corresponding gauge transformation $\Omega_1(x)$ is not uniquely defined.
Any $SU(N-1)$ transformation among the components $2,3,..,N$ can still be
performed. This is consistent with the number of remaining degrees of freedom:
$(2 N - 1)$ constraints must be satisfied to bring $v_1(x)$ parallel to a
reference direction, which leaves $(N^2 - 1) - (2 N - 1) = (N-1)^2 - 1$ d.o.f., 
corresponding to an $SU(N-1)$ gauge freedom. 
Choose any satisfactory $\Omega_1$ and rotate all eigenvectors 
$v_1, v_2,..,v_{N-1}$. \\
$ii)$ Rotate the last $(N-1)$ components of $v_2(x)$ to make it parallel
to $(1,0,..,0)_{N-1}^\dagger$, leaving the first component untouched.
This defines a gauge transformation $\Omega_2(x) \in {\bf 1} \times SU(N-1)$,
up to a further $SU(N-2)$ transformation. 
Choose any satisfactory $\Omega_2$ and rotate all eigenvectors. \\
$iii)$ Rotate the last $(N-2)$ components of $v_3(x)$ parallel to
$(1,0,..,0)_{N-2}^\dagger$, etc... \\
$iv)$ Finally the last 2 components of $v_{N-1}(x)$ are rotated parallel
to $(1,0)^\dagger$, defining a gauge transformation 
$\Omega_{N-1}(x) \in {\bf 1}_{N-2} \times SU(2)$, as in Section 2, and 
completing the gauge fixing.

The complete gauge transformation is therefore 
$\Omega_{N-1} \times ... \times \Omega_2 \times \Omega_1$.
It turns the matrix $M$ formed by $v_1,v_2,..,v_N$ to upper triangular form,
regardless of the value of $v_N$, which needs not be computed.
Our procedure can be viewed as a $QR$ decomposition of $M$.

Let us now consider the ambiguities associated with our $SU(N)$ procedure. \\
$\bullet$ Each eigenvector is defined up to a complex phase, which is global 
and will induce an irrelevant, global Abelian gauge transformation. \\
$\bullet$ The gauge becomes ambiguous when any of the lowest $(N-1)$ 
eigenvalues is degenerate, preventing a unique determination (up to a phase)
of the eigenvectors. This defines a genuine Gribov copy. In this case,
a continuous family of gauges is obtained, corresponding to different choices
of orthogonal eigenvectors in the degenerate subspace. \\
$\bullet$ On a generic field configuration, {\em local} gauge ambiguities
arise at points $x$ where all the last $(N-k+1)$ complex components of $v_k(x)$
vanish, making the gauge transformation $\Omega_k(x)$ ill-defined.
This implies $2(N-k+1)$ real constraints, thus defining
defects of co-dimension $2(N-k+1)$. 

These defects all have a topological meaning. For example, $SU(5)$ will admit
point-like defects in $d=10$ dimensions ($N=5,k=1$), 
consistent with the non-trivial homotopy group $\Pi_{2N-1}(SU(N)) = Z$.
Even though such excitations are irrelevant classically because, in $d > 4$,
their action decreases with their size, their role could be studied 
non-perturbatively via Laplacian gauge-fixing.

In 4 dimensions, the only defects of co-dimension $\le 4$ are point-like,
and come from the last step in our gauge-fixing procedure, which involves
an $SU(2)$ gauge transformation. Thus, one recovers the expected result
that the topological charge of an $SU(N)$ configuration can be ``squeezed''
to an $SU(2)$ subgroup [even if the instantons themselves may not be],
where it gives rise to local singularities in a smooth gauge. Again, this 
offers the possibility of extracting the topological charge of an $SU(N)$
configuration without cooling.

\vspace{0.3cm}
{\bf Acknowledgments:}
We thank C.~Alexandrou, J.L.F.~Barb{\'o}n, F.~Bruckmann, J.~Fr\"ohlich, 
M.~Garc{\'\i}a~P\'erez, M.~L\"uscher and A.~Wipf for correspondence and 
discussions.

\end{document}